# Efficient Web Log Mining using Doubly Linked Tree

Ratnesh Kumar Jain[1] , Dr. R. S. Kasana[1]

[1]Department of Computer Science & Applications,
Dr. H. S. Gour, University,
Sagar, MP (India)
jratnesh@rediffmail.com, irkasana7158@gmail.com

Dr. Suresh Jain[2]

[2]Department of Computer Engineering, Institute of
Engineering & Technology,
Devi Ahilya University,
Indore, MP (India)
suresh.jain@rediffmail.com

**Abstract— World Wide Web is a huge data repository and is growing with the explosive rate of about 1 million pages a day. As the information available on World Wide Web is growing the usage of the web sites is also growing. Web log records each access of the web page and number of entries in the web logs is increasing rapidly. These web logs, when mined properly can provide useful information for decision-making. The designer of the web site, analyst and management executives are interested in extracting this hidden information from web logs for decision making. Web access pattern, which is the frequently used sequence of accesses, is one of the important information that can be mined from the web logs. This information can be used to gather business intelligence to improve sales and advertisement, personalization for a user, to analyze system performance and to improve the web site organization. There exist many techniques to mine access patterns from the web logs. This paper describes the powerful algorithm that mines the web logs efficiently. Proposed algorithm firstly converts the web access data available in a special doubly linked tree. Each access is called an event. This tree keeps the critical mining related information in very compressed form based on the frequent event count. Proposed recursive algorithm uses this tree to efficiently find all access patterns that satisfy user specified criteria. To prove that our algorithm is efficient from the other GSP (Generalized Sequential Pattern) algorithms we have done experimental studies on sample data.**

*Keywords: Web mining; Pattern discovery.*

## I. INTRODUCTION

The World Wide Web provides almost unlimited access to the documents on the Internet to its users. Web mining is a specialized field of data mining. In web mining we apply data mining techniques on the huge repository of web data. Web mining can be categorized into web content mining, web structure mining and web usage mining. Web usage mining looks at the log of Web access. Web server records each access of the web page in web logs. Number of entries in the web logs is increasing rapidly as the access to the web site is increasing. These web logs, when mined properly can provide useful information for decision-making. Most of the web logs contain information about fields: IP Address, User Name, Time Stamp, Access Request, Result Status, Byte Transferred,

Referrer URL and User Agent. There are many efforts towards mining various patterns from Web logs [4, 9,11].

Web access patterns mined from Web logs can be used for purposes like: Improving design of web sites, used to gather business intelligence to improve sales and advertisement, analyzing system performance, building adaptive Web sites [7, 6, 10].

Finding access pattern is the problem of finding association rules in the Web logs [2]. The problem of finding association rules falls within the purview of database mining [1,5] also called knowledge discovery in databases. Mining frequent access patterns (called sequential access pattern mining) in a sequence database was firstly introduced by Agrawal and Srikant [3] which is based on AprioriAll algorithm. After its introduction lots of work was done to mine sequential pattern efficiently. Srikant and Agrawal in 1996 [8] gave a generalized sequential pattern mining algorithm, GSP, which outperforms their AprioriAll algorithm. In this algorithm sequence database is scanned many times to mine sequential access pattern. In the first scan, it finds all frequent 1-event and forms a set of 1-event frequent sequences. In the following scans, it generates candidate sequences from the set of frequent sequences and checks their supports. The problem with GSP is that it does not perform well if the length of the access sequences and transactions are large, which is the basic need of Web log mining.

In this paper we address the problem of handling large access patterns efficiently. Our solution is consists of two phases: In the first phase we compress the presentation of access sequences using doubly linked tree structure and in the second phase we apply the mining algorithm to efficiently mine all the frequent access sequences. We give a performance study in support of our work, which proves that this mining algorithm is faster than the other Apriori-based GSP mining algorithms.

## II. PROBLEM STATEMENTS

As we have discussed a Web log consist of many types of information including the information about the user and the access done by the user. We can extract the unnecessary data and only keep the required data in the preprocessing phase of the log mining. If each access is regarded as event we can say that after preprocessing web log is a sequence of events from one user or session in timestamp ascending order. Let us define some terms before finally stating the problem.

Let E be a set of events. Then $S = e_1 e_2 \cdots e_k e_{k+1} \cdots e_n$ ($e_i \in$ E) for ($1 \le i \le n$) is an access sequence and n is the length of the access sequence called n-sequence. Remember in any access sequence repetition is allowed i.e. $e_i \neq e_j$ for $i \neq j$. Access sequence $S' = e_1' e_2' e_3' \ldots e_k'$ is called a subsequence of sequence S and S is called the super-sequence of sequence S' denoted as $S' \subseteq S$, if and only if there exist $1 \le i_1 < i_2 < \cdots \ i_k \le n$, such that $e_j' = e_{i_j}$ for ($1 \le j \le k$). S' is called proper subsequence of sequence S that is $S' \subset S$, if and only if S' is a subsequence of S and S' $\neq$ S. Subsequence $S_s = e_{k+1} e_{k+2} \cdots e_n$ of S is a super-sequence of a sequence $P = e_{k+1} e_{k+2} \cdots e_l$ where $l \le n$ then $S_p = e_1 e_2 \cdots e_k$ is called the prefix of S with respect to sequence P and $S_s$ is called the suffix of $S_p$.

Let Web access sequence database WAS is represented as a set $\{S_1, S_2 \cdots, S_m\}$ where each $S_i$ ($1 \le i \le m$) are access sequences. Then the support of access sequence S in WAS is defined as $Sup(S) = \frac{\left| S_i \mid S \subseteq S_i \right|}{m}$. A sequence S is said a $\xi$-pattern or simply (Web) access pattern of WAS, if $Sup(s) \ge \xi$. Here it is important to remember that events can be repeated in an access sequence or pattern, and any pattern can get support at most one from one access sequence.

**The problem of mining access pattern is**: Given Web access sequence database WAS and a support threshold $\xi$, mine the complete set of $\xi$-pattern of WAS.

### III. Efficient Web Log Mining using Doubly Linked Tree

Most of the previously proposed methods were based on the **Apriori heuristic**. According to Apriori "if a sequence G is not a $\xi$-pattern of sequence database, any super-sequence of G cannot be a $\xi$-pattern of sequence database."

Though this property may substantially reduce the size of candidate sets but the combinatorial nature of the pattern mining, it may still generate a huge set of candidate patterns, especially when the sequential pattern is long. This motivates us to introduce some new technique for Web access pattern mining. The central theme of our algorithm is as follows:

Scan the WAS twice. In the first scan, determine the set of frequent events. An event is called a frequent event if and only if it appears in at least ($\xi$ . |WAS| ). Where |WAS| denotes the number of access sequences in WAS and $\xi$ denote the support threshold. In the second scan, build a doubly linked tree After creating a doubly linked tree we recursively mine it using conditional search to find all $\xi$-pattern.

The following observations are helpful in the construction of the doubly linked tree.

1. Apriori property that if a sequence G is not a $\xi$-pattern of sequence database, any super-sequence of G cannot be a $\xi$-pattern of sequence database is used. That means, if an event e is not in the set of frequent 1-sequences, there is no need to include e in the construction of a doubly linked tree.

2. We create a single branch for the shared prefix P in the tree. It helps in saving space and support counting of any subsequence of the prefix P.

Above observation suggest that doubly linked tree should be defined to contain following information:

- Each node must contain event (we call it label) and its count except the root node which have empty label and count=0. The count specifies the number of occurrences of the corresponding prefix ended with that event in the WAS.

- To manage the linkage and backward traversal we need two additional pointers except the pointers tree normally has. First, all the nodes in the tree with the same label are linked by a queue called event-node queue. To maintain the front of a queue for each frequent event in the tree one header table is maintained. Second, for backward traversal from any intermediate node to the root we add a pointer to the parent at each node.

The tree construction process is as follows: First of all filter out the nonfrequent events from each access sequence in WAS and then insert the resulting frequent subsequence into tree started from the root. Considering the first event, denoted as e, increment the count of child node with label e by 1 if there exists one; otherwise create a child labeled by e and set the count to 1. Then, recursively insert the rest of the frequent subsequence to the subtree rooted at that child labeled e. The complete algorithm for doubly linked tree creation is given below:

**Algorithm 1 (Doubly Linked Tree Construction)**

Input: A Web access sequence database WAS and a set of all possible events E.

Output: A doubly linked tree T.

Method:

**Scan 1:**
1. For each access sequence S of the WAS
   1.1. For each event in E
      1.1.1. For each event of an access sequence of WAS. If selected event of access sequence is equal to selected event of E then
         a. event count = event count + 1
         b. continue with the next event in E.
2. For each event in E if event qualify the threshold add that event in the set of frequent event **FE**.

**Scan 2:**

1. Create a root node for T

2. For each access sequence S in the access sequence database WAS do

    (a) Extract frequent subsequence S' from S by removing all events appearing in S but not in FE. Let $S' = s_1 s_2 \cdots s_n$, where $s_i$ ($1 \le i \le n$) are events in S'. Let current node is a pointer that is currently pointing to the root of T.

    (b) For i=1 to n do, if current node has a child labeled $s_i$, increase the count of $s_i$ by 1 and make current node point to $s_i$, else create a new child node with label= $s_i$, count =1, parent pointer = current node and make current node point to the new node, and insert it into the $s_i$ -queue.

3. Return (T);

After the execution of this algorithm we get doubly linked tree. This contains all the information in very condensed form. Now we do not need WAS database to mine the access pattern. The length of the tree is one plus the maximum length of the frequent subsequences in the database. The width of the tree that is the number of distinct leaf nodes as well as paths in a doubly linked tree cannot be more than the number of distinct frequent subsequences in the WAS database. Access sequences with same prefix will share some upper part of path from root and due to this scheme size of the tree is much smaller than the size of WAS database.

Maintaining some additional links provides some interesting properties which helps in mining frequent access sequences.

    1. For any frequent event $e_i$, all the frequent subsequences contain $e_i$ can be visited by following the $e_i$ -queue, starting from the record for $e_i$ in the header table of doubly linked tree.

    2. For any node labeled $e_i$ in a doubly linked tree, all nodes in the path from root of the tree to this node (excluded) form a prefix sequence of $e_i$. The count of this node labeled $e_i$ is called the count of the prefix sequence.

    3. A path from root may have more than one node labeled $e_i$, thus a prefix sequence say G of $e_i$ if it contain another prefix sequence say H of $e_i$ then G is called the **super-prefix sequence** and H is called the **sub-prefix sequence**. The problem is that super-prefix sequence contributes in the counting of sub-prefix sequence. This problem is resolved using **unsubsumed count.** A prefix sequence of $e_i$ without any super-prefix sequences, unsubsumed count is the count of $e_i$. For a prefix sequence of $e_i$ with some super-prefix sequences, the unsubsumed count of it is the count of that sequence minus unsubsumed counts of all its super-prefix sequences.

4. It is very difficult to traverse from root to the node pointed by the $e_i$ -queue because it requires several traversal hits to get required prefix. Parent pointer allows **backward traversal** from any intermediate node pointed by $e_i$ -queue to the root and efficiently extract the prefix sequences.

With the above information we can apply **conditional search** to mine all Web access patterns using doubly linked tree. Conditional search means, instead of searching all Web access patterns at a time, it turns to search Web access patterns with same **suffix**. This suffix is then used as the condition to narrow the search space. As the suffix becomes longer, the remaining search space becomes smaller potentially. The algorithm to mine all ξ-patterns is as follows:

**Algorithm 2 (Mining all ξ-patterns in doubly linked tree)**

Input: a Doubly linked tree **T** and support threshold ξ.

Output: the complete set of ξ-patterns.

Method:

    1. If doubly linked tree T has only one branch, return all the unique combinations of nodes in that branch

    2. Initialize Web access pattern set WAP= $\phi$. Every event in T itself is a Web access pattern, insert them into WAP

    3. For each event $e_i$ in T,

        a. Construct a conditional sequence base of $e_i$, i.e. PS( $e_i$ ), by following the $e_i$ -queue, count conditional frequent events at the same time

        b. If the set of conditional frequent events is not empty, build a conditional doubly linked tree for $e_i$ over PS( $e_i$ ) using algorithm 1. Recursively mine the conditional doubly linked tree

        c. For each Web access pattern returned from mining the conditional doubly linked tree, concatenate $e_i$ to it and insert it into WAP

    4. Return WAP.

## IV. PERFORMANCE EVALUATION

To compare the performance of Doubly Linked Tree mine Algorithm and GSP Algorithm, both are implemented with C++ language running under Turbo C++. All experiments are performed on 2.16GHz NoteBook Computer with 1GB of RAM. The Operating System is Microsoft XP Professional Version 2002 Service Pack 2. To illustrate the performance comparisons we used freely available web access logs on internet. Since these web logs are in different format we did some preprocessing work to convert this weblog into the Web Access Pattern Dataset (WASD) format. This original Web logs are recorded in server from 23 feb, 2004 to 29 feb, 2004 are 61 KB in size, and 36,878 entries which includes 7851 unique host names, 2793 unique URLs and 32 different file types (extensions). We filter out the web logs according to our need.

There are so many measures to compare efficiency of the two algorithms. We here used run time as a measure of efficiency. To compare the performance of Doubly Linked Tree mine and GSP, we did several experiments that can be categorized into two types. In the first type of experiments we checked the performance with respect to the threshold for a fixed size of WASD. In the second type of experiments we checked the performance with respect to the size of the WASD for a fixed support threshold.

As the results shows performance of the doubly linked tree mining out performs GSP in both the cases. As can be seen in figures when the support threshold is low Doubly Linked Tree mining algorithm took approx. 100 sec. while GSP took 450 sec. and this difference reduces as support threshold increases. And as the size of data base increases, run time of GSP increases more rapidly than Doubly Linked Tree mining.

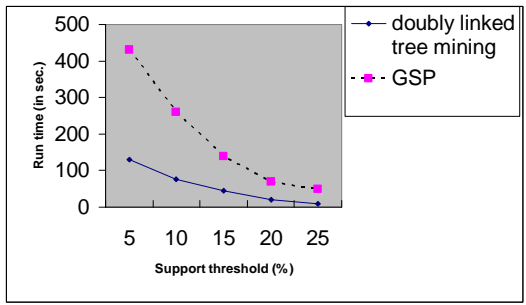

(a) Comparison for varying support threshold

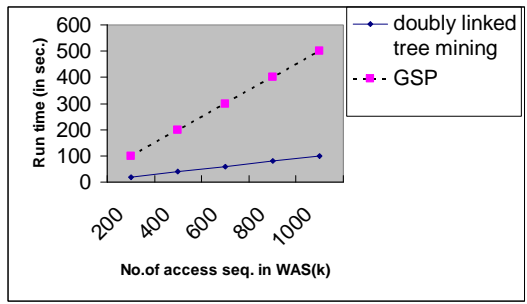

(b) Comparison for varying Number of Access Sequences

**Figure 1: Experimental Results of the Comparative study between Doubly Linked Tree Mining and GSP Algorithms.**

## V. CONCLUSION

As shown in figure 1(a) and 1(b) we can say that run time required by GSP for any support threshold and for any size of Web Access Sequence Database is higher than Doubly Linked Tree mining Algorithm. For low support threshold and for large data base Doubly Linked Tree mining performance is much better than GSP. While for higher support threshold and small size of data set since only few events qualify the criteria of frequent event there is no significant difference in both the algorithms. The comparison proves that Doubly Linked Tree mining Algorithm is more efficient than GSP especially for low support threshold and large Web Access Sequence Database.
For mining sequential patterns from web logs, the following aspects may be considered for **future work**. Some algorithm should be developed so that we do not need to do preprocessing work manually, rather these mining algorithms can be applied directly on the web log files. Also efficient web usage mining could benefit from relating usage of the web page to the content of web page. Some other area of interest may be implementing Doubly Linked mine algorithm to the distributed environment

### ACKNOWLEDGMENT

Author is grateful to the technical reviewers for the comments, which improved the clarity and presentation of the paper. Author wishes to thank Dr. Pankaj Chaturvedi and Mr. Deepak Sahu for all the discussions and contributions during the initial stages of this research.

### AUTHORS PROFILE

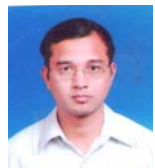

**Ratnesh Kumar Jain** is currently a research scholar at Department of Computer Science and Applications, Dr. H. S. Gour Central University (formerly, Sagar University) Sagar, M P, India. He completed his bachelor's degree in Science (B. Sc.) with Electronics as special subject in 1998 and master's degree in computer applications (M.C.A.) in 2001 from the same University. His field of study is Operating System, Data Structures, Web mining, and Information retrieval. He has published more than 4 research papers and has authored a book.

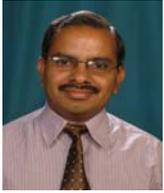

**Suresh Jain** completed his bachelor's degree in civil engineering from Maulana Azad National Institute of Technology (MANIT) (formerly, Maulana Azad College of Technology) Bhopal, M.P., India in 1986. He completed his master's degree in computer engineering from S.G. Institute of Technology and Science, Indore in 1988, and doctoral studies (Ph.D. in computer science) from Devi Ahilya University, Indore. He is professor of Computer Engineering in Institute of Engineering & Technology (IET), Devi Ahilya University, Indore. He has experience of over 21 years in the field of academics and research. His field of study is grammatical inference, machine learning, web mining, and information retrieval. He has published more than 25 research papers and has authored a book.

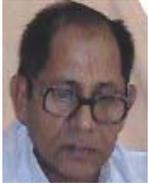

**R. S. Kasana** completed his bacholar's degree in 1969 from Meerut University, Meerut, UP, India. He completed his master's degree in Science (M.Sc.-Physics) and master's degree in technology (M. Tech.-Applied Optics) from I.I.T. New Delhi, India. He completed his doctoral and post doctoral studies from Ujjain University in 1976 in Physics and from P. T. B. Braunschweig and Berlin, Germany & R.D. Univ. Jabalpur correspondingly. He is a senior Professor and HoD of Computer Science and Applications Department of Dr. H. S. Gour University, Sagar, M P, India. During his tenure he has worked as vice chancellor, Dean of Science Faculty, Chairman Board of studies. He has more than 34 years of experience in the field of academics and research. Twelve Ph. D. has awarded under his supervision and more than 110 research articles/papers has published.